\documentstyle[amsfonts,epsfig,12pt]{article}

\def\mypagenumber{1}

\def\myend{\end{document}}

\normalsize

\newcounter{sxn}

\newcounter{axn}

\date{}

\newdimen\mybaselineskip
\newcommand{\beeq}{\begin{equation}}
\newcommand{\eneq}{\end{equation}}
\newcommand{\be}{\begin{eqnarray}}
\newcommand{\ee}{\end{eqnarray}}
\newcommand{\bpic}{\begin{picture}}
\newcommand{\epic}{\end{picture}}

\def\la{\raise.16ex\hbox{$\langle$} \, }
\def\ra{\, \raise.16ex\hbox{$\rangle$} }

\def\psibar{ \psi \kern-.65em\raise.6em\hbox{$-$} }
\def\mbar{ m \kern-.78em\raise.4em\hbox{$-$}\lower.4em\hbox{} }

\def\n@space{\nulldelimiterspace=0pt \mathsurround=0pt }
\def\huge#1{{\hbox{$\left#1\vbox to 20.5pt{}\right.\n@space$}}}

\def\myskip{\noalign{\kern 8pt}}
\def\myeqspace{\noalign{\kern 10pt}}

\def\boxit#1{$\vcenter{\hrule\hbox{\vrule\kern3pt
    \vbox{\kern3pt\hbox{#1}\kern3pt}\kern3pt\vrule}\hrule}$}
\def\bigbox#1{$\vcenter{\hrule\hbox{\vrule\kern5pt
     \vbox{\kern5pt\hbox{#1}\kern5pt}\kern5pt\vrule}\hrule}$}

\def\ignore#1{{}}

\begin{document}

\bibliographystyle{unsrt}
\footskip 1.0cm

\thispagestyle{empty}
\setcounter{page}{\mypagenumber}

%{\baselineskip=10pt \parindent=0pt \small
%\mydate 
%}                
             
\begin{flushright}{
BRX-TH-495\\}

\end{flushright}

\vspace{2.5cm}
\begin{center}
{\LARGE \bf {Newtonian Counterparts of Spin 2 Massless Discontinuities}}\\ 
\vskip 1 cm
{\large{S. Deser and Bayram  Tekin  }}\footnote{e-mail:~
deser, tekin@brandeis.edu}\\
\vspace{.5cm}
{\it Department of  Physics, Brandeis University, Waltham,02454  MA,
USA}\\

\end{center}

\vspace*{2.5cm}
%\baselinestretch{2.0}

%\normalsize

\begin{abstract}
\baselineskip=18pt
Massive spin 2 theories in flat or cosmological ( $\Lambda \ne 0$)
backgrounds are subject to discontinuities as the masses tend to zero.
We show and explain physically why their Newtonian limits do not inherit this behaviour.
On the other hand, conventional ``Newtonian cosmology'' , where $\Lambda $ is a 
constant source of the potential, displays discontinuities: {\it{e.g.}} for any finite range,
$\Lambda$ can be totally removed.

\end{abstract}
\vfill

%\end{titlepage}
 
\newpage

%\setcounter{page}{1}

%\textheight=20cm
%\headsep=0.75cm
%\vsize=20cm

%%%%%%%%%%%%%%%%%%%%%%%%%%%%%%%%%%%%%%%%%%%%%%%%%%%%%%%%%%%%%%%%%
\normalsize
\baselineskip=22pt plus 1pt minus 1pt
\parindent=25pt
\vskip 2 cm

It is well known that higher spin fields in flat space lead to finitely different interactions
among their prescribed, conserved~\footnote{The massless, gauge, theories are consistent  
only if the sources are fixed and conserved.}, sources depending on whether they are strictly massless 
or have a mass, however small. 
This possible discontinuity, absent for spins less than $3/2$, is universal
for higher spins. It was first found explicitly for spin $2$   \cite{vandam,zakharov} and  for spin $3/2$  
 \cite{deser1}.
More recently \cite{higuchi,kogan,porrati,deser2} the question  has been re-opened for these models 
when they propagate in a background cosmological 
($\Lambda \ne 0$ ) space. 
The presence of this second dimensional constant provides alternative
paths, and outcomes, for the massless limit. In particular, the spin 2 case with, say, two (background 
covariantly conserved )  sources
$ ( T_{\mu\nu}, t^{\mu \nu}) $ leads to the Born exchange interaction,
\be
I = G_{\Lambda,m} \int d^4 x \left \{  T_{\mu \nu}\, {\cal{D}}\, t^{\mu\nu} - 
{ m^2 - \Lambda \over 3m^2 - 2 \Lambda} T_{\mu}\,^\mu \, {\cal{D}}\, t_{\nu}\,^\nu \right \},
\label{interaction}
\ee
where $ {\cal{D}}$ is the usual massive $(A)dS$ scalar propagator whose $m=0$ and $\Lambda =0$ limits are smooth   
and $G_{\Lambda,m}$ is the gravitational  constant for the particular $(\Lambda,m)$ model. The old
discontinuity~\footnote{The effect of $1/3$ versus $1/2$ was a
finite discrepancy between predictions for experiments involving only slow 
( $t_{\mu\nu} \rightarrow t_{00}$ only) and those involving light-like 
( {\it{e.g.}} $t_\mu\,^\mu = 0 $) sources. For,
and  only for, the value $1/2$ could both light bending and 
Newtonian gravity agree with observation since 
the coupling constant $G_{\Lambda, m}$ is used up to fix the latter's strength.} 
 at $\Lambda =0$ led to a relative coefficient $1/3$ in the second term versus $1/2$
if $m^2$ is identically zero. 
When $\Lambda \ne 0$, there is an infinite number of limits available; in particular 
$m^2 \rightarrow 0$ followed by  $\Lambda \rightarrow 0$ reproduces the 1/2 factor.
The fermionic spin $3/2$ case is similar but with additional subtleties 
\cite{andrew1,andrew} concerning the meaning
of ``masslessness'' when $\Lambda \ne 0$.
Our purpose here is to
discuss the same set of problems in the Newtonian counterparts of the above linearised models
as well as in traditional Newtonian cosmology. We do not consider here non-linear massive
gravity because it is neither viable \cite{deser3} nor perturbatively linearisable~\footnote{
While it may be possible to obtain a massless limit to the ( non-linear !) Schwarzschild metric 
\cite{vainshtein, vainshtein2} and then linearise, this would constitute a very different ``Newtonian
limit ''. } \cite{vainshtein,jun}.

Before considering the details, we argue physically that the Newtonian limit of (\ref{interaction})
must be immune to discontinuities because by its very definition, it is only valid for 
$c \rightarrow \infty$. Thus {\it{only}} ( $ T_0\,^0 = \rho, t_0\,^0 = \sigma$ ) fail to vanish: 
we have an effective scalar theory with only slow sources and one ``experiment'' to fit with 
one coupling constant. There is no ``light-bending'' to fit, as there is no light ($c = \infty$).  

If $\Lambda = 0$, the interaction is
\be
I_{0,m} \sim {2\over 3}G_{0,m}\, \int d^3 x \,\rho\, Y\, \sigma \, ,
\label{yukawa}
\ee
where $Y$ is the Yukawa potential and 
$2 G_{0,m}/3$ is tuned to the observed Newtonian constant. 
Since the Yukawa potential reduces continuously to  $1/r$, the
$m \rightarrow 0$ process is  perfectly smooth. 

If, on the other hand, $\Lambda \ne 0$, the effective interaction becomes 
\be
I_{\Lambda,m} \sim G_{\Lambda,m}\, ( 1- {m^2 -\Lambda \over 3 m^2 - 2\Lambda} )\, \int d^3 x \,\rho\, Y_\Lambda\, \sigma\, , 
\ee
where $Y_\Lambda$ is the generalized static Yukawa potential when $\Lambda \ne 0$. Thus 
``Newton's constant'' is
\be
G_{N} =G_{\Lambda,m}\, {2 m^2- \Lambda \over 3 m^2 - 2\Lambda }.  
\label{fraction}
\ee

This $(m^2,\Lambda)$ dependence of $G_{N}$ would seem to involve some dangerous ranges and points. However,
in the original theory whose limit this is, all models with $ 0 < 3 m^2 < 2 \Lambda $ are non-unitary and so
unphysical \cite{higuchi,andrew}; for us, this excludes the region where the fraction in (\ref{fraction}) would turn
negative, as well as the point $2m^2 = \Lambda$ where the numerator vanishes~\footnote{We emphasize that, at the Newtonian
level, this vanishing is a simple case of cancellation between Newtonian attraction and the non-unitary helicity zero ghost's
repulsive contribution. It corresponds to the covariant interaction $ (T_{\mu\nu}t^{\mu\nu}- 
T_{\mu}^{\mu}t_{\nu}^{\nu}) $ in which there is manifestly no $T_{00}t^{00}$ term.}. 
The $3m^2 = 2\Lambda$ model \cite{nepo} is unitary but has a gauge invariance
that requires its conserved sources to be traceless as well, so it has no Newtonian limit at all.
The physical region relevant to (\ref{fraction}) thus consists of the usual gauge point $m^2 = 0$, 
together with that part of the ($m^2, \Lambda$)
plane for which $m^2 > 2\Lambda/3$, including of course $AdS$ space where $\Lambda < 0$. Any limit of  
$(m^2, \Lambda )\rightarrow 0$ in this region is perfectly smooth, with a well-defined positive 
$G_{N}$.  

We now turn to a different, if similarly named, model,
Newtonian cosmology (see for example \cite{trautman}). This is neither the above Newtonian 
limit of linearised gravity about its 
$(A)dS$ vacuum backgrounds, nor even obviously that about the (false) flat vacuum. 
It consists of a Poisson equation, with constant background, for the potential
\be
\nabla^2 \Phi  - \Lambda = 0,
\ee
whose homogeneous solution is 
\be
\Phi_{\Lambda, 0} =  C + {\Lambda r^2 \over 6}
\label{lambda1}
\ee
Throughout we omit (for brevity only) all localized sources.
Adding a finite range would then lead to the most general non-relativistic system 
~\footnote{For a nice historical account of finite range Newtonian forces
(in the absence of a cosmological term ), first studied by Neumann and Seeliger
in the late 19th century, as well as Einstein's ideas on the Newtonian limit 
of $\Lambda \ne 0$ General Relativity, we refer the reader to \cite{norton}. }

\be
\nabla^2 \Phi - m^2 \Phi - \Lambda = 0,
\ee
 whose generic homogeneous solution is 
\be
\Phi_{\Lambda,m}= - {\Lambda \over m^2} + D ({ e^{m r}\over r} -  {e^{-m r}\over r} )\, .  
\label{generic}
\ee
The integration constant $(C, D)$ above are arbitrary. We had to include the rising ``anti-Yukawa'' 
exponential to avoid a (localised source-like) singularity at the origin. Normally, if $\Lambda =0$, we would
immediately set $D = 0$ to retain acceptable behaviour of $\Phi$ at infinity. By contrast, while
the massless solution (\ref{lambda1}) also rises at infinity, this is necessity: $\Lambda$ is {\it{not}}
an integration constant. Let us now follow the consequences of the two options for $D$:

 1.  If the physical choice $ D= 0$ is taken in (\ref{generic}), so that $\Phi_{\Lambda,m}= - \Lambda/  m^2$, then
an obvious -constant- shift of $\Phi$ removes all traces of $\Lambda$. But this in turn leads to a different, truly
cosmic $m \rightarrow 0 $ discontinuity: we have lost the essential $\Lambda r^2/6$ term in (\ref{lambda1}) altogether.
 
2.  Keeping $D \ne 0$, does allow a smooth massless limit if one also tunes it to be 
$D =  \Lambda / 2 m^3$. With this choice one indeed recovers (\ref{lambda1}) for
$m \rightarrow 0$. [ Even the irrelevant constant $C$ in (\ref{lambda1}) can be reproduced
by adding $C/2m$ to $D$, at the cost of bad behaviour at $\Lambda =0$: $\Phi_{0,m}$
blows up exponentially. ] Therefore one can only recover 
both $\Lambda \rightarrow 0$ and $m \rightarrow 0$ limits smoothly with the above choice of $D$ at an 
unacceptably high price: not only did we need the exponentially rising solutions but also the integration
constant $D$ had to be tuned to the parameters $(\Lambda, m)$ of the model.

Our study of Newtonian limits has borne out the physical argument that a theory with a single source
($T_{00}$) and a single scalar field has no scope for ``interesting'' behaviour. We showed that
all {\it{unitary}}  massive spin 2 theories coupled to conserved, traceful, $T_{\mu \nu}$ have Newtonian limits
smooth in  $(m^2, \Lambda )$. Instead, Newtonian cosmology depends on $(m^2, \Lambda)$ in ways that do permit 
qualitative discontinuities, as exemplified by the fact that any  $(m^2 \ne 0, \Lambda)$ model is equivalent
to one with $(m^2, 0)$, but does {\it{not}} limit to the  $(0, \Lambda)$ ones.

We thank A. Waldron, and the authors of \cite{vainshtein2}, for stimulating correspondence.

This work was supported by National Science Foundation  grant PHY99-73935

\myend